\documentclass[aps,pre,subeqn,nofootinbib,amsmath,amsfonts]{revtex4}
\usepackage{hyperref}
\usepackage{graphicx, xcolor}
\usepackage{longtable}
\usepackage{subfigure}
\usepackage{euscript}

\newcommand{\eqr}[1]{Eq.~\eqref{#1}}

\newcommand{\secr}[1]{Sec.~[\ref{#1}]}

\newcommand{\tblr}[1]{Table~[\ref{#1}]}

\newcommand{\appr}[1]{Appendix~[\ref{#1}]}

\newcommand{\velocity}{{\mathrm{v}}}
\newcommand{\vvelocity}{{\mathbf{v}}}
\newcommand{\vJ}{{\mathbf{J}}}
\newcommand{\vR}{{\mathbf{r}}}

\newcommand{\vg}{{\mathbf{g}}}

\newcommand{\nmorange}{\overline{1,n\!-\!1}}

\newcommand{\spd}{\!\cdot\!}
\newcommand{\xs}{x^{\scriptstyle s}}

\newcommand{\xgsb}{x^{g,s}}
\newcommand{\xlsb}{x^{\ell,s}}

\newcommand{\lame}{\mathfrak{h}}

\graphicspath{{Figures/}}

\begin{document}
\title{Integral relations for the surface transfer coefficients.}
\author{K.~S.~Glavatskiy$^{1}$}
\author{D.~Bedeaux$^{1,2}$}
\affiliation
{%
$^{1}$Department of Chemistry, Norwegian University of Science and Technology, NO 7491 Trondheim, Norway.\\
$^{2}$Department of Process and Energy, Technical University of Delft, Leeghwaterstr 44, 2628 CA Delft, The Netherlands.
}%
\date\today
\begin{abstract}
In this paper we derive relations between the local resistivities inside the interfacial region and the overall resistances of the surface between two phases for a mixture. These resistivities
are the coefficients in the force-flux relations for the stationary heat and mass transfer through the interface. We have shown that interfacial resistances depend among other things on the
enthalpy profile across the interface. Since this variation is substantial (the enthalpy of evaporation is one of the main differences between liquid and vapor phases) the interfacial
resistivities are also substantial. Particularly, surface put up much more resistance to the heat and mass transfer then the homogeneous phase. This is the case not only for the pure heat
conduction and diffusion but also for the cross effects like thermal diffusion.

\end{abstract}
\maketitle
%
%\tableofcontents
%
\numberwithin{equation}{section}
\section{Introduction}
%

%$\mathfrak{ABCDEFGHIJKLMNOPQRSTUVWXYZ}$
%$\mathfrak{abcdefghijklmnopqrstuvwxyz}$
%$\mathcal{ABCDEFGHIJKLMNOPQRSTUVWXYZ}$
%$\mathscr{ABCDEFGHIJKLMNOPQRSTUVWXYZ}$

During evaporation and condensation heat and mass is transported through the interface. The common description of these phenomena uses certain assumptions which are debatable. For instance, one
usually assumes equilibrium conditions at the interface \cite{Krishna/MassTransfer} and neglects the coupling effects between the heat and mass transfer \cite{kjelstrupbedeaux/heterogeneous}.
The first one is a zero approximation and it is desirable to extend it to a more accurate theory. This is done in \cite{kjelstrupbedeaux/heterogeneous} for instance. Neglecting the coupling
coefficients was shown to violate the second law of thermodynamics for a one-component system \cite{Bedeaux2005}. The coupling is important since the corresponding transfer coefficients depend
on the enthalpy of vaporization. The significance of this quantity in this context is due to the large difference between the liquid and the vapor values of the enthalpy.

The situation becomes even more complicated when one consider evaporation in mixtures. These processes happen in such industrial applications as distillation and therefore the precise
description is important. Depending on conditions one can get the mass fluxes of components in the same or in the opposite directions.

There has been done a number of studies of the interfacial transport for one-component systems: experiments \cite{Fang1999a, Bedeaux1999b, Mills2002, James2006}, molecular dynamic simulations
\cite{Rosjorde2000, Rosjorde2001, Kjelstrup2002, Simon2004, jialin/longrange} , kinetic theory \cite{Pao1971a, Sone1973, Cipolla1974, Bedeaux1990} and square gradient continuous description
\cite{bedeaux/vdW/III, bedeaux/vdW/IV}. All these works use different approaches, which allows one to investigate different aspects. Mainly one-component systems have been studied. One of the
points of interest is the dependence of the overall interfacial resistances on the continuous profiles. Once we have a description which relates the resistance of the Gibbs surface
\cite{Gibbs/ScientificPapers} to the continuous profiles of, in particular, local resistivities, we can describe the surface separately from the adjacent bulk phases which is closely related to
hypothesis of local equilibrium of a surface \cite{kjelstrupbedeaux/heterogeneous}.

In this paper we extend the analysis done in \cite{bedeaux/vdW/IV} for one-component system. Here we do the analysis for mixtures. Using the square gradient theory for the non-equilibrium
interface developed earlier \cite{glav/grad1} we derive the general relations for the case of the transport perpendicular to the surface. We show that, as in the case of a one-component system,
one can obtain the interfacial resistances using the continuous profiles obtained for equilibrium. This simplifies analysis a lot since one does not need to consider a non-equilibrium solution
in order to obtain these resistivities. This is in fact a requirement, which the interfacial resistances must satisfy: as being defined within the linear force-flux relations they should depend
only on unperturbed quantities, i.e. equilibrium ones.

The evaporation and condensation often take place not only through planar interfaces, but also through curved ones, like the evaporation into a bubble. We do not restrict ourselves to the planar
interfaces and give the expressions for the interfacial resistivities for curved surfaces.

In \cite{glav/grad3} we have obtained the overall interfacial resistivities $R^{g}$ and $R_{\ell}$ using three different methods: an experiment-like procedure, a perturbation cell method and
kinetic theory (only $R^{g}$). Those methods were found to be in a good agreement. In this paper we compare them with the interfacial resistivities found using integral relations, which relates
these quantities to the local resistivities directly. We show that the agreement is also good.

In this paper we will focus on the heat and mass transfer through the interface of a two-phase mixture. We will not consider non-equilibrium perturbation along the surface. We will also assume
the fluid to be non-viscous. Furthermore we will focus on the stationary non-equilibrium perturbation. In \secr{sec/EntropyProduction} we give the expressions for the local and the excess
entropy production found in \cite{glav/grad1} and \cite{glav/grad3}. In \secr{sec/Resistivities} we derive the integral relations in general form. It is convenient to use measurable heat and
mass fluxes and we therefore show how to translate general relations to the resistivities associated with the measurable fluxes in \secr{sec/Measurable}. In \secr{sec/Binary} we give the
explicit expressions for a binary mixture and apply the analysis to the particular mixture of cyclohexane and $n$-hexane. Concluding remarks are given in \secr{sec/Discussion}.

\section{The entropy production.}\label{sec/EntropyProduction}
\subsection{Stationary states.}\label{sec/EntropyProduction/Stationary}

Consider the total energy flux $\vJ_{e}$ and the mass fluxes $\vJ_{\xi_{i}} \equiv \rho_{i}\,\vvelocity_{i}$ and $\vJ_{m} \equiv \rho\,\vvelocity$, where $\rho_{i}$ and $\rho$ are the density of
$i$-th component and the overall density respectively, while $\vvelocity_{i}$ and $\vvelocity \equiv \sum_{i=1}^{n-1}{\xi_{i}\vvelocity_{i}}$ are the velocity of the $i$-th component and the
barycentric velocity respectively. Furthermore $\xi_{i}$ is the mass fraction of $i$-th component and $n$ is the total number of components. In the stationary states these fluxes satisfy the
relations
\begin{equation}\label{eq/ExcessEntropy/01}
\nabla\spd\vJ_{e} = 0 ,\quad \nabla\spd\vJ_{\xi_{i}} = 0 ,\quad \nabla\spd\vJ_{m} = 0
\end{equation}%
As we are interested in transport through the surface we will restrict analysis to solutions of the form $\vJ(x_{1}, \mathbf{x}_{\parallel}) = (J(x_{1}), 0, 0)$, where $\vJ$ is one of the above
fluxes. Furthermore $x_{1}$ is the normal coordinate to the surface and $\mathbf{x}_{\parallel} \equiv (x_{2}, x_{3})$ are the tangential coordinates. In the case of normal transport all the
quantities depend only on $x_{1}$ but not on $\mathbf{x}_{\parallel}$. For these solutions
\begin{equation}\label{eq/ExcessEntropy/01a}
\nabla\spd\vJ = \nabla_{\perp}\,J =  \frac{1}{\lame_{1}\lame_{2}\lame_{3}}\frac{d }{d x_{1}}(\lame_{2}\,\lame_{3}\,J) = 0
\end{equation}%
where $\lame_{i} \equiv \lame_{i}(x_{1}, \mathbf{x}_{\parallel})$ are Lame coefficients for curvilinear coordinates. It follows from \eqr{eq/ExcessEntropy/01a} that
\begin{equation}\label{eq/ExcessEntropy/01b}
\lame_{2}(x_{1})\,\lame_{3}(x_{1})\,J(x_{1}) = \lame_{2}(x^{s})\,\lame_{3}(x^{s})\,J(x^{s}) \equiv \lame_{2}^{s}\,\lame_{3}^{s}\,J^{s}
\end{equation}%
where $x^{s}$ is the chosen dividing surface and $\lame_{2}^{s}$ and $\lame_{3}^{s}$ are the Lame coefficients on that dividing surface. We will suppress the superscript $s$ for the flux $J$ as
long as it does not lead to confusion.

\subsection{Local entropy production.}\label{sec/EntropyProduction/Local}

Consider the local entropy production for the mixture interface found in \cite{glav/grad1}, written for the case of transport in the direction through the interface

\begin{equation}\label{eq/ExcessEntropy/02}
\sigma_{s} = \displaystyle J_{q}\,\nabla_{\perp}\frac{1}{T} - \sum_{i=1}^{n-1}{J_{i}\,\nabla_{\perp}\frac{\psi_{i}}{T}} = J_{q}\,\nabla_{\perp}\frac{1}{T} -
\sum_{i=1}^{n}{J_{i}\,\nabla_{\perp}\frac{\mu_{i}}{T}}
\end{equation}%
where $T$ is the temperature, $J_{q}$ is the heat flux and $J_{i} \equiv \rho_{i}\,(\velocity_{i}-\velocity) = J_{\xi_{i}} - \xi_{i}\,J_{m}$ is the diffusive mass flux, which satisfies the
relation
\begin{equation}\label{eq/ExcessEntropy/02a}
\sum_{i=1}^{n}{J_{i}} = 0
\end{equation}%
Furthermore $\psi_{i} = \mu_{i}-\mu_{n}$, where $\mu_{i}$ is the chemical potential of the $i$-th component. The heat flux is related to the total energy flux and to the measurable heat flux
$J_{q}'$ as
\begin{equation}\label{eq/ExcessEntropy/03}
J_{q} = J_{e} - J_{m}\,(h + \velocity^{2}/2-\vg\spd\vR) = J_{q}^{\,\prime} + \sum_{i=1}^{n}{h_{i}J_{i}}
\end{equation}%
Here $h_{i}$ and $h$ are the partial enthalpy of the $i$-th component and the specific enthalpy respectively, while $\vg$ is the gravitational acceleration and $\velocity^{2}/2-\vg\spd\vR$ is
the sum of specific kinetic and potential energies of the mixture.

For the stationary transport through the interface the Gibbs-Duhem relation has the following form (see \cite{glav/grad3} for details):
\begin{equation}\label{eq/ExcessEntropy/04}
s\,\frac{\partial T}{\partial x_{1}} + \sum_{i=1}^{n}{\xi_{i}\frac{\partial \mu_{i}}{\partial x_{1}}} - v\,\frac{\partial \sigma_{11}}{\partial x_{1}} = 0
\end{equation}%
where $\sigma_{11}$ is an element of the thermodynamic pressure tensor $\sigma_{\alpha\beta}$. Together with the equation of motion, $\nabla_{\perp}\sigma_{11} +
\rho\,\nabla_{\perp}(\velocity^{2}/2-\vg\spd\vR) = 0$, it gives
\begin{equation}\label{eq/ExcessEntropy/04a}
\sum_{i=1}^{n}{\xi_{i}\left(\nabla\frac{\widetilde{\mu}_{i}}{T} - \widetilde{h}_{i}\nabla\frac{1}{T}\right)} = 0
\end{equation}%
where $\tilde{\mu}_{i} \equiv \mu_{i} + \velocity^{2}/2-\vg\spd\vR$ and $\tilde{h}_{i} \equiv h_{i} + \velocity^{2}/2-\vg\spd\vR$. Substituting \eqr{eq/ExcessEntropy/03} and $J_{i}$ into
\eqr{eq/ExcessEntropy/02} and using \eqr{eq/ExcessEntropy/04a} one can express the local entropy production \eqref{eq/ExcessEntropy/02} in terms of the total energy flux:
\begin{equation}\label{eq/ExcessEntropy/05}
\sigma_{s} = J_{e}\,\nabla_{\perp}\frac{1}{T} - \sum_{i=1}^{n}{J_{\xi_{i}}\,\nabla_{\perp}\frac{\tilde{\mu}_{i}}{T}}
\end{equation}%
and in terms of the measurable heat flux:
\begin{equation}\label{eq/ExcessEntropy/06}
\sigma_{s} = J_{q}^{\,\prime}\,\nabla_{\perp}\frac{1}{T} - \sum_{i=1}^{n-1}{J_{i}\,\left(\nabla_{\perp}\frac{\psi_{i}}{T} - \eta_{i}\,\nabla_{\perp}\frac{1}{T}\right)}\\
\end{equation}%
where $\eta_{i} \equiv \widetilde{h}_{i}-\widetilde{h}_{n} = h_{i}-h_{n}$. Note that the difference between partial enthalpies with a tilde is equal to the difference without the tilde. In
stationary states \eqr{eq/ExcessEntropy/05}, \eqr{eq/ExcessEntropy/06} and \eqr{eq/ExcessEntropy/02} are completely equivalent expressions.

\subsection{Excess entropy production.}\label{sec/EntropyProduction/Excess}

We define the excess $\widehat{\phi}(\xs)$ of a density $\phi(\xs)$ per unit of volume in curvilinear coordinates as
\begin{equation}\label{eq/ExcessEntropy/07}
\widehat{\phi}(\xs) \equiv \frac{1}{\lame_{2}^{s}\,\lame_{3}^{s}}\,\int_{\displaystyle \xgsb}^{\displaystyle \xlsb}{dx_{1}\,\lame_{1}\,\lame_{2}\,\lame_{3}\,\phi^{ex}(x_{1}; \xs)}
\end{equation}
where
\begin{equation}\label{eq/ExcessEntropy/08}
\phi^{ex}(x_{1}; \xs) \equiv \phi(x_{1}) - \phi^{g}(x_{1})\,\Theta(\xs-x_{1}) - \phi^{\ell}(x_{1})\,\Theta(x_{1}-\xs)
\end{equation}
Here $\xs$ indicates the position of the chosen dividing surface while $\xgsb$ and $\xlsb$ are the boundaries of the interfacial region at the gas and liquid side respectively. These boundaries
are chosen such that $\phi(\xgsb) = \phi^{g}(\xgsb)$ and $\phi(\xlsb) = \phi^{\ell}(\xlsb)$ with a certain accuracy. Superscripts $\ell$ and $g$ indicate the function $\phi$ extrapolated from
the liquid and gas to the surface region. We refer to \cite{glav/grad3} for more extensive discussion.

Taking the excess of the local entropy production given by \eqr{eq/ExcessEntropy/05} we obtain (see \cite{glav/grad3} for details):
\begin{equation}\label{eq/ExcessEntropy/09}
\widehat{\sigma}_{s} = J_{e}\left(\frac{1}{T^{\ell}} - \frac{1}{T^{g}}\right) -  \sum_{i=1}^{n}{J_{\xi_{i}}\left(\frac{\tilde{\mu}_{i}^{\ell}}{T^{\ell}}-
\frac{\tilde{\mu}_{i}^{g}}{T^{g}}\right)}
\end{equation}%
Here $T^{\ell} \equiv T^{\ell}(\xs)$ and $T^{g} \equiv T^{g}(\xs)$ are the temperatures extrapolated from the liquid and gas to the dividing surface. The analogous meaning have
$\tilde{\mu}_{i}^{\ell}$ and $\tilde{\mu}_{i}^{g}$. All the quantities, both the fluxes and the forces, are evaluated at the dividing surface $\xs$.

Using \eqr{eq/ExcessEntropy/03} we obtain
\begin{equation}\label{eq/ExcessEntropy/10}
\widehat{\sigma}_{s} = J_{q}^{\,\prime,\,g}\left(\frac{1}{T^{\ell}} - \frac{1}{T^{g}}\right) -  \sum_{i=1}^{n}{J_{\xi_{i}}\left[\left(\frac{\tilde{\mu}_{i}^{\ell}}{T^{\ell}} -
\frac{\tilde{\mu}_{i}^{g}}{T^{g}}\right) - \tilde{h}_{i}^{g}\left(\frac{1}{T^{\ell}}-\frac{1}{T^{g}}\right) \right]}
\end{equation}%
and
\begin{equation}\label{eq/ExcessEntropy/11}
\widehat{\sigma}_{s} = J_{q}^{\,\prime,\,\ell}\left(\frac{1}{T^{\ell}} - \frac{1}{T^{g}}\right) -  \sum_{i=1}^{n}{J_{\xi_{i}}\left[\left(\frac{\tilde{\mu}_{i}^{\ell}}{T^{\ell}} -
\frac{\tilde{\mu}_{i}^{g}}{T^{g}}\right) - \tilde{h}_{i}^{\ell}\left(\frac{1}{T^{\ell}}-\frac{1}{T^{g}}\right) \right]}
\end{equation}%
where $J_{q}^{\,\prime,\,\ell}$ and $J_{q}^{\,\prime,\,g}$ are the measurable heat fluxes on the liquid and gas side respectively. Again \eqr{eq/ExcessEntropy/10}, \eqr{eq/ExcessEntropy/11} and
\eqr{eq/ExcessEntropy/09} are completely equivalent in stationary states.

\section{Integral relations.}\label{sec/Resistivities}

Consider the entropy production \eqr{eq/ExcessEntropy/05}, each term in which has a form $J\,\nabla_{\perp}\phi$. Following the common procedure in non-equilibrium thermodynamics one can write
the force-flux relations for those entropy productions. Since all the terms have the same form, it is sufficient to consider only one force-flux pair. The phenomenological relation for that pair
then reads
\begin{equation}\label{eq/Resistivities/11}
\nabla_{\perp}\phi(x_{1}) \equiv \frac{1}{\lame_{1}}\frac{d \phi(x_{1})}{d x_{1}} = r(x_{1})\,J(x_{1})
\end{equation}%
for the local entropy production. The corresponding term in the excess entropy production in \eqr{eq/ExcessEntropy/09} has a form $J(\phi^{\ell}-\phi^{g})$. The phenomenological relation
relation for this term reads
\begin{equation}\label{eq/Resistivities/12}
\phi^{\ell}(\xs)-\phi^{g}(\xs) = R(\xs)\,J(\xs)
\end{equation}%
For the general case $\phi$ and $J$ must be replaced by a set as well as $R$ and $r$ by the corresponding matrix.

Let us introduce excess operators $\mathfrak{E}_{n}$ and $\mathfrak{E}_{r}$ which we will apply to a quantity $\phi$. Let
\begin{equation}\label{eq/Resistivities/03}
\mathfrak{E}_{n}\{\phi\} \equiv \displaystyle \int_{\displaystyle \xgsb}^{\displaystyle \xlsb}{dx_{1}\lame_{1}\,\phi^{ex}(\vR; \xs)}
\end{equation}%
and
\begin{equation}\label{eq/Resistivities/04}
\mathfrak{E}_{r}\{\phi\} \equiv \lame_{2}^{s}\,\lame_{3}^{s} \int_{\displaystyle \xgsb}^{\displaystyle \xlsb}{dx_{1}\frac{\lame_{1}}{\lame_{2}\lame_{3}}\,\phi^{ex}(\vR; \xs)}
\end{equation}%
where $\phi^{ex}$ is defined by \eqr{eq/ExcessEntropy/08}. In cartesian coordinates the excess operators $\mathfrak{E}_{n}$, $\mathfrak{E}_{r}$ and the excess $\widehat{\hphantom{\phi}}$, are
given be the same expression when $\phi$ is a density per unit of volume. One should not confuse them however, as these operators have different meanings. The excess $\widehat{\phi}$ may be
applied only to a volume density $\phi$ and means the surface density. In contrast, neither $\mathfrak{E}_{n}\{\phi\}$ nor $\mathfrak{E}_{r}\{\phi\}$ need to be applied to a volume density. In
\eqr{eq/ExcessEntropy/05} $\phi$ can be the inverse temperature or a chemical potential divided by the temperature.

Applying $\mathfrak{E}_{n}$ operator to the both sides of \eqr{eq/Resistivities/11} one can show that
\begin{equation}\label{eq/Resistivities/13}
\phi^{\ell}(\xs) - \phi^{g}(\xs) = \mathfrak{E}_{r}\{r\}\,J(\xs)
\end{equation}%
Comparing \eqr{eq/Resistivities/13} with \eqr{eq/Resistivities/12} we conclude that
\begin{equation}\label{eq/Resistivities/14}
R = \mathfrak{E}_{r}\{r\}
\end{equation}%
This is the general form of the integral relation for the resistivities $r$ and $R$. \eqr{eq/Resistivities/13} together with \eqr{eq/Resistivities/14} are the most important and fundamental
results of the paper. The generalization to a set of $\phi$ and $J$ is straightforward and will be applied in the rest of the paper.

We then proceed to the explicit expressions for the phenological equations. The local entropy production \eqref{eq/ExcessEntropy/05} produces the following force-flux relations
\begin{equation}\label{eq/Resistivities/01}
\begin{array}{rl}
\displaystyle \nabla_{\perp}\frac{1}{T} &= \displaystyle r^{e}_{qq}\,J_{e} - \sum_{i=1}^{n}{r^{e}_{qi}\,J_{\xi_{i}}} \\
\displaystyle \nabla_{\perp}\frac{\tilde{\mu}_{j}}{T} &= \displaystyle r^{e}_{jq}\,J_{e} - \sum_{i=1}^{n}{r^{e}_{ji}\,J_{\xi_{i}}}
\end{array}
\end{equation}%
and the excess entropy production \eqref{eq/ExcessEntropy/09} produces
\begin{equation}\label{eq/Resistivities/02}
\begin{array}{rl}
\displaystyle \frac{1}{T^{\ell}} - \frac{1}{T^{g}} &= \displaystyle R^{e}_{qq}\,J_{e} -  \sum_{i=1}^{n}{R^{e}_{qi}\,J_{\xi_{i}}} \\
\displaystyle \frac{\tilde{\mu}_{j}^{\ell}}{T^{\ell}}- \frac{\tilde{\mu}_{j}^{g}}{T^{g}} &= \displaystyle R^{e}_{jq}\,J_{e} - \sum_{i=1}^{n}{R^{e}_{ji}\,J_{\xi_{i}}}
\end{array}
\end{equation}%
where $J_{e}$ and $J_{\xi_{i}}$ as well as $T$ and $\mu_{j}$ are evaluated at the dividing surface $\xs$. The off-diagonal coefficients of both sets satisfy the Onsager reciprocal relations.

Comparing \eqr{eq/Resistivities/01} with \eqr{eq/Resistivities/02} and using \eqr{eq/Resistivities/13} together with \eqr{eq/Resistivities/14} we may conclude that
\begin{equation}\label{eq/Resistivities/06}
\begin{array}{rl}
R^{e}_{qq} &= \mathfrak{E}_{r}\,\{r^{e}_{qq}\} \\
R^{e}_{qi} &= \mathfrak{E}_{r}\,\{r^{e}_{qi}\} = \mathfrak{E}_{r}\,\{r^{e}_{iq}\} = R^{e}_{iq}  \\
R^{e}_{ji} &= \mathfrak{E}_{r}\,\{r^{e}_{ji}\} = \mathfrak{E}_{r}\,\{r^{e}_{ij}\} = R^{e}_{ij}  \\
\end{array}
\end{equation}%

\eqr{eq/Resistivities/06} represents integral relations for the resistivity coefficients associated with the total energy flux.

\section{Measurable heat fluxes.}\label{sec/Measurable}

It is convenient to obtain the integral relations for the interface resistivities associated with the measurable heat fluxes rather then the total energy flux. We obtain the relations between
resistivities, both local and for the whole surface.

First a note regarding the dependence of the resistivities on the reference state of for instance the enthalpy. Both $R^{e}$ and $r^{e}$ coefficients depend on the reference state, as they are
the resistivities associated with the absolute fluxes. The resistivities $R^{\,\prime\,g}$ and $r^{\,\prime}$, which will be defined below, are associated with the measurable fluxes and
therefore independent of the reference state.

\subsection{The whole surface.}\label{sec/Measurable/Global}

In order to obtain the resistivities for the whole surface we consider \eqr{eq/ExcessEntropy/10} for the entropy production in terms of the measurable heat flux on the gas side of the surface.
The analysis for the liquid side is completely equivalent to the one done for the gas side.

The entropy production \eqref{eq/ExcessEntropy/10} produces the following phenomenological equations
\begin{equation}\label{eq/Measurable/01}
\begin{array}{rl}
\displaystyle \frac{1}{T^{\ell}} - \frac{1}{T^{g}} &= \displaystyle R^{\,\prime\,g}_{qq}\,J_{q}^{\,\prime,\,g} -  \sum_{i=1}^{n}{R^{\,\prime\,g}_{qi}\,J_{\xi_{i}}} \\
\displaystyle \left(\frac{\tilde{\mu}_{i}^{\ell}}{T^{\ell}} - \frac{\tilde{\mu}_{i}^{g}}{T^{g}}\right) - \tilde{h}_{i}^{g}\left(\frac{1}{T^{\ell}}-\frac{1}{T^{g}}\right) %
&= \displaystyle R^{\,\prime\,g}_{jq}\,J_{q}^{\,\prime,\,g} - \sum_{i=1}^{n}{R^{\,\prime\,g}_{ji}\,J_{\xi_{i}}}
\end{array}
\end{equation}%

The resistivities from \eqr{eq/Measurable/01} and \eqr{eq/Resistivities/02} are related, using \eqr{eq/ExcessEntropy/03} by
\begin{equation}\label{eq/Measurable/02}
\begin{array}{rl}
R^{e}_{qq} &= R^{\,\prime\,g}_{qq}\\\\
R^{e}_{qi} &= R^{\,\prime\,g}_{qi} + \widetilde{h}_{i}^{g}\,R^{\,\prime\,g}_{qq} \\\\
R^{e}_{ji} &= R^{\,\prime\,g}_{ji} + \widetilde{h}_{i}^{g}\,R^{\,\prime\,g}_{jq} + \widetilde{h}_{j}^{g}\,R^{\,\prime\,g}_{qi} + \widetilde{h}_{i}^{g}\,\widetilde{h}_{j}^{g}\,R^{\,\prime\,g}_{qq}\\\\
\end{array}
\end{equation}
\eqr{eq/Measurable/01} and \eqr{eq/Resistivities/02} are \textit{linear} relations between the forces and the fluxes. In \eqr{eq/Measurable/02} we should therefore use coexistence values of the
enthalpies.

%According to linear theory, all the resistivities should be determined only by equilibrium non-perturbed properties. Thus we should write $\widetilde{h}_{i,\,eq}$ instead of $\widetilde{h}_{i}$.
%With no loss of generality the former quantity is equal to simply $h_{i,\,eq}$ (the gravitational term may be eliminated by either the proper choice of the coordinate origin or by making it a
%part of $h_{i,\,eq}$).
Inverting \eqr{eq/Measurable/02} and using the equilibrium enthalpies we obtain for the resistivities associated with the measurable heat flux
\begin{equation}\label{eq/Measurable/03}
\begin{array}{rl}
R^{\,\prime\,g}_{qq} &= R^{e}_{qq}\\\\
R^{\,\prime\,g}_{qi} &= R^{e}_{qi} - \widetilde{h}_{i,\,eq}^{g}\,R^{e}_{qq} \\\\
R^{\,\prime\,g}_{ji} &= R^{e}_{ji} - \widetilde{h}_{i,\,eq}^{g}\,R^{e}_{jq} - \widetilde{h}_{j,\,eq}^{g}\,R^{e}_{qi} + \widetilde{h}_{i,\,eq}^{g}\,\widetilde{h}_{j,\,eq}^{g}\,R^{e}_{qq}\\\\
\end{array}
\end{equation}
where $\widetilde{h}_{i,\,eq} = {h}_{i,\,eq}(\xs) - \vg\spd\vR^{s}$ and all the quantities are evaluated at the chosen dividing surface.

\subsection{Local resistivities.}\label{sec/Measurable/Local}

In order to write the linear laws for the local forces and fluxes we use \eqr{eq/ExcessEntropy/06} and \eqr{eq/ExcessEntropy/02}. The entropy production \eqref{eq/ExcessEntropy/06} gives

\begin{equation}\label{eq/Measurable/04}
\begin{array}{rl}
\displaystyle \nabla_{\perp}\frac{1}{T} &= \displaystyle r^{\,\prime}_{qq}\,J_{q}^{\,\prime} -  \sum_{i=1}^{n-1}{r^{\,\prime}_{qi}\,J_{i}} \\
\displaystyle \nabla_{\perp}\frac{\psi_{i}}{T} - \eta_{i}\nabla_{\perp}\frac{1}{T} &= \displaystyle r^{\,\prime}_{jq}\,J_{q}^{\,\prime} - \sum_{i=1}^{n-1}{r^{\,\prime}_{ji}\,J_{i}}
\end{array}
\end{equation}
while the entropy production \eqref{eq/ExcessEntropy/02} gives
\begin{equation}\label{eq/Measurable/05}
\begin{array}{rl}
\displaystyle \nabla_{\perp}\frac{1}{T} &= \displaystyle r_{qq}\,J_{q} -  \sum_{i=1}^{n-1}{r_{qi}\,J_{i}} \\
\displaystyle \nabla_{\perp}\frac{\psi_{i}}{T} &= \displaystyle r_{jq}\,J_{q} - \sum_{i=1}^{n-1}{r_{ji}\,J_{i}}
\end{array}
\end{equation}
The local resistivities $r$ and $r^{\,\prime}$ are related as follows
\begin{equation}\label{eq/Measurable/045}
\begin{array}{rl}
r^{\,\prime}_{qq} &= r_{qq}\\
r^{\,\prime}_{qi} &= r_{qi} - \eta_{i}\,r_{qq}\\
r^{\,\prime}_{ji} &= r_{ji} - \eta_{i}\,r_{jq} - \eta_{j}\,r_{qi} + \eta_{i}\eta_{j}\,r_{qq}
\end{array}
\end{equation}
Again, we should use the equilibrium profiles for $\eta_{i}$.

We now relate the local resistivities associated with the measurable heat flux to the local resistivities associated with the total energy flux, similarly\footnote{The details of this procedure
are given in \appr{sec/Appendix/Resistivities}.} to how it was done for the overall surface resistivities. Comparing \eqr{eq/Measurable/04} with \eqr{eq/Resistivities/01} we obtain
\begin{equation}\label{eq/Measurable/06}
\begin{array}{rl}
r^{e}_{qq} =& \displaystyle r^{\,\prime}_{qq} \\
r^{e}_{qi} =& \displaystyle r^{\,\prime}_{qq}\widetilde{h}_{i} - \sum_{k=1}^{n-1}{r^{\,\prime}_{qk}\,\xi_{k}} + r^{\,\prime}_{qi},\quad i=\nmorange \\
r^{e}_{en} =& \displaystyle r^{\,\prime}_{qq}\widetilde{h}_{n} - \sum_{k=1}^{n-1}{r^{\,\prime}_{qk}\,\xi_{k}} \\
r^{e}_{ji} =& \displaystyle r^{\,\prime}_{qq}\widetilde{h}_{j}\widetilde{h}_{i} - \sum_{k=1}^{n-1}{\xi_{k}(r^{\,\prime}_{ke}\widetilde{h}_{i}+r^{\,\prime}_{qk}\widetilde{h}_{j})} + r^{\,\prime}_{jq}\widetilde{h}_{i}+r^{\,\prime}_{qi}\widetilde{h}_{j}+ \\
        & \displaystyle + r^{\,\prime}_{ji} - \sum_{k=1}^{n-1}{\xi_{k}(r^{\,\prime}_{ki}+r^{\,\prime}_{jk})} + \sum_{k=1}^{n-1}\sum_{l=1}^{n-1}{r^{\,\prime}_{kl}\,\xi_{k}\,\xi_{l}},\quad j,i=\nmorange \\
r^{e}_{ni} =& \displaystyle r^{\,\prime}_{qq}\widetilde{h}_{n}\widetilde{h}_{i} - \sum_{k=1}^{n-1}{\xi_{k}(r^{\,\prime}_{ke}\widetilde{h}_{i}+r^{\,\prime}_{qk}\widetilde{h}_{n})} + r^{\,\prime}_{qi}\widetilde{h}_{n} - \sum_{k=1}^{n-1}{r^{\,\prime}_{ki}\,\xi_{k}} + \sum_{k=1}^{n-1}\sum_{l=1}^{n-1}{r^{\,\prime}_{kl}\,\xi_{k}\,\xi_{l}},\quad i=\nmorange \\
r^{e}_{nn} =& \displaystyle r^{\,\prime}_{qq}\widetilde{h}_{n}^{2} - \widetilde{h}_{n}\sum_{k=1}^{n-1}{\xi_{k}(r^{\,\prime}_{ke}+r^{\,\prime}_{qk})} + \sum_{k=1}^{n-1}\sum_{l=1}^{n-1}{r^{\,\prime}_{kl}\,\xi_{k}\,\xi_{l}} \\
\end{array}
\end{equation}
and comparing \eqr{eq/Measurable/05} with \eqr{eq/Resistivities/01} we obtain
\begin{equation}\label{eq/Measurable/07}
\begin{array}{rl}
r^{e}_{qq} =& \displaystyle r_{qq} \\
r^{e}_{qi} =& \displaystyle r_{qq}\widetilde{h} - \sum_{k=1}^{n-1}{r_{qk}\,\xi_{k}} + r_{qi},\quad i=\nmorange \\
r^{e}_{en} =& \displaystyle r_{qq}\widetilde{h} - \sum_{k=1}^{n-1}{r_{qk}\,\xi_{k}} \\
r^{e}_{ji} =& \displaystyle r_{qq}\widetilde{h}^{2} - \widetilde{h}\sum_{k=1}^{n-1}{\xi_{k}(r_{ke}+r_{qk})} + \widetilde{h}(r_{jq}+r_{qi})+ \\
        & \displaystyle + r_{ji} - \sum_{k=1}^{n-1}{\xi_{k}(r_{ki}+r_{jk})} + \sum_{k=1}^{n-1}\sum_{l=1}^{n-1}{r_{kl}\,\xi_{k}\,\xi_{l}},\quad j,i=\nmorange \\
r^{e}_{ni} =& \displaystyle r_{qq}\widetilde{h}^{2} - \widetilde{h}\sum_{k=1}^{n-1}{\xi_{k}(r_{ke}+r_{qk})} + \widetilde{h}r_{qi} - \sum_{k=1}^{n-1}{r_{ki}\,\xi_{k}} + \sum_{k=1}^{n-1}\sum_{l=1}^{n-1}{r_{kl}\,\xi_{k}\,\xi_{l}},\quad i=\nmorange \\
r^{e}_{nn} =& \displaystyle r_{qq}\widetilde{h}^{2} - \widetilde{h}\sum_{k=1}^{n-1}{\xi_{k}(r_{ke}+r_{qk})} + \sum_{k=1}^{n-1}\sum_{l=1}^{n-1}{r_{kl}\,\xi_{k}\,\xi_{l}} \\
\end{array}
\end{equation}
where $\widetilde{h} \equiv h + \velocity^{2}/2-\vg\spd\vR$.

Again, as in \eqr{eq/Measurable/02} we should use the enthalpy and the mass fraction profiles in \eqr{eq/Measurable/06} and \eqr{eq/Measurable/07} in equilibrium. This leads to the relation
$\widetilde{h}_{i} = h_{i,\,eq}-\vg\spd\vR$ and $\widetilde{h} = h_{eq}-\vg\spd\vR$.

\section{Results for the two component mixture.}\label{sec/Binary}

In this section we derive the integral relations for a binary mixture. The formula given in this section is not restricted to a particular binary mixture however. We do that for the sake of
convenience as well as because we will apply them to a particular two component mixture. For simplicity of notation we will not write subscript $_{eq}$ for enthalpies, keeping in mind that all
of them should be evaluated at equilibrium.

The relations \eqref{eq/Measurable/06}-\eqref{eq/Measurable/07} between resistivities in case of binary mixture take the following form
\begin{equation}\label{eq/Binary/01}
\begin{array}{rl}
r^{e}_{qq} =& r^{\,\prime}_{qq} \\
r^{e}_{q1} =& r^{\,\prime}_{qq}\,\widetilde{h}_{1} + r^{\,\prime}_{q1}\,\xi_{2} \\
r^{e}_{q2} =& r^{\,\prime}_{qq}\,\widetilde{h}_{2} - r^{\,\prime}_{q1}\,\xi_{1} \\
r^{e}_{11} =& r^{\,\prime}_{qq}\,\widetilde{h}_{1}^{2} +2r^{\,\prime}_{q1}\,\xi_{2}\,\widetilde{h}_{1} + r^{\,\prime}_{11}\,\xi_{2}^{2} \\
r^{e}_{21} =& r^{\,\prime}_{qq}\,\widetilde{h}_{2}\,\widetilde{h}_{1}  + r^{\,\prime}_{q1}(\xi_{2}\,\widetilde{h}_{2} - \xi_{1}\,\widetilde{h}_{1}) - r^{\,\prime}_{11}\,\xi_{1}\,\xi_{2} \\
r^{e}_{22} =& r^{\,\prime}_{qq}\,\widetilde{h}_{2}^{2} - 2r^{\,\prime}_{q1}\,\xi_{1}\,\widetilde{h}_{2} + r^{\,\prime}_{11}\,\xi_{1}^{2} \\
\end{array}
\end{equation}
and
\begin{equation}\label{eq/Binary/02}
\begin{array}{rl}
r^{e}_{qq} =& r_{qq} \\
r^{e}_{q1} =& r_{qq}\,\widetilde{h} + r_{q1}\,\xi_{2}\\
r^{e}_{q2} =& r_{qq}\,\widetilde{h} - r_{q1}\,\xi_{1} \\
r^{e}_{11} =& r_{qq}\,\widetilde{h}^{2} + 2r_{q1}\,\xi_{2}\,\widetilde{h} + r_{11}\,\xi_{2}^{2} \\
r^{e}_{21} =& r_{qq}\,\widetilde{h}^{2} + r_{q1}(\xi_{2} - \xi_{1})\widetilde{h} - r_{11}\,\xi_{1}\,\xi_{2} \\
r^{e}_{22} =& r_{qq}\,\widetilde{h}^{2} - 2r_{q1}\,\xi_{1}\,\widetilde{h} + r_{11}\,\xi_{1}^{2} \\
\end{array}
\end{equation}
respectively. Using \eqr{eq/Resistivities/06} and \eqr{eq/Measurable/03} we therefore obtain
\begin{equation}\label{eq/Binary/03}
\begin{array}{rl}
R^{\,\prime\,g}_{qq} =& \mathfrak{E}_{r}\,\{r^{\,\prime}_{qq}\} \\\\
R^{\,\prime\,g}_{q1} =& \mathfrak{E}_{r}\,\{r^{\,\prime}_{qq}(h_{1}-h_{1}^{g}) + r^{\,\prime}_{q1}\,\xi_{2} \} \\\\
R^{\,\prime\,g}_{q2} =& \mathfrak{E}_{r}\,\{r^{\,\prime}_{qq}(h_{2}-h_{2}^{g}) - r^{\,\prime}_{q1}\,\xi_{1} \} \\\\
R^{\,\prime\,g}_{11} =& \mathfrak{E}_{r}\,\{r^{\,\prime}_{qq}(h_{1}-h_{1}^{g})^{2} + 2r^{\,\prime}_{q1}\,\xi_{2}\,(h_{1}-h_{1}^{g}) + r^{\,\prime}_{11}\,\xi_{2}^{2} \} \\\\
R^{\,\prime\,g}_{12} =& \mathfrak{E}_{r}\,\{r^{\,\prime}_{qq}(h_{1}-h_{1}^{g})(h_{2}-h_{2}^{g}) + r^{\,\prime}_{q1}(\xi_{2}\,(h_{2}-h_{2}^{g}) - \xi_{1}\,(h_{1}-h_{1}^{g})) - r^{\,\prime}_{11}\,\xi_{1}\,\xi_{2} \} \\\\
R^{\,\prime\,g}_{22} =& \mathfrak{E}_{r}\,\{r^{\,\prime}_{qq}(h_{2}-h_{2}^{g})^{2} - 2r^{\,\prime}_{q1}\,\xi_{1}\,(h_{2}-h_{2}^{g}) + r^{\,\prime}_{11}\,\xi_{1}^{2} \} \\\\
\end{array}
\end{equation}
and
\begin{equation}\label{eq/Binary/04}
\begin{array}{rl}
R^{\,\prime\,g}_{qq} =& \mathfrak{E}_{r}\,\{r_{qq}\} \\\\
R^{\,\prime\,g}_{q1} =& \mathfrak{E}_{r}\,\{r_{qq}(h-h_{1}^{g}) + r_{q1}\,\xi_{2} \} \\\\
R^{\,\prime\,g}_{q2} =& \mathfrak{E}_{r}\,\{r_{qq}(h-h_{2}^{g}) - r_{q1}\,\xi_{1} \} \\\\
R^{\,\prime\,g}_{11} =& \mathfrak{E}_{r}\,\{r_{qq}(h-h_{1}^{g})^{2} + 2r_{q1}\,\xi_{2}\,(h-h_{1}^{g}) + r_{11}\,\xi_{2}^{2} \} \\\\
R^{\,\prime\,g}_{12} =& \mathfrak{E}_{r}\,\{r_{qq}(h-h_{1}^{g})(h-h_{2}^{g}) + r_{q1}(\xi_{2}\,(h-h_{2}^{g}) - \xi_{1}\,(h-h_{1}^{g})) - r_{11}\,\xi_{1}\,\xi_{2} \} \\\\
R^{\,\prime\,g}_{22} =& \mathfrak{E}_{r}\,\{r_{qq}(h-h_{2}^{g})^{2} - 2r_{q1}\,\xi_{1}\,(h-h_{2}^{g}) + r_{11}\,\xi_{1}^{2} \} \\\\
\end{array}
\end{equation}
Furthermore we used that $\widetilde{h}_{i} - \widetilde{h}_{i}^{g} = h_{i}-h_{i}^{g}$ in both equations. \eqr{eq/Binary/03} and \eqr{eq/Binary/04} are the integral relations for the
resistivities associated with the measurable heat flux for a binary mixture. They clearly do not depend on the reference chosen for the enthalpies.

We further need the values for the local resistivities. In \cite{glav/grad2} we developed a scheme which uses $r$ coefficients in calculation. We therefore use the following expressions
\begin{equation}\label{eq/Binary/05}
\begin{array}{rl}
r_{qq}(x) &= r_{qq}^{g} + (r_{qq}^{\ell}-r_{qq}^{g})\,q_{0}(x) + \alpha_{qq}(r_{qq}^{\ell}+r_{qq}^{g})\,q_{1}(x)\\
\\
r_{q1}(x) &= r_{q1}^{g} + (r_{q1}^{\ell}-r_{q1}^{g})\,q_{0}(x) + \alpha_{q1}(r_{q1}^{\ell}+r_{q1}^{g})\,q_{1}(x)\\
\\
r_{11}(x) &= r_{11}^{g} + (r_{11}^{\ell}-r_{11}^{g})\,q_{0}(x) + \alpha_{11}(r_{11}^{\ell}+r_{11}^{g})\,q_{1}(x)\\
\end{array}
\end{equation}
where $q_{0}(x)$ and $q_{1}(x)$ for each resistivity are modulatory curves for the resistivity profiles: $q_{0}(x)$ is a smooth $\arctan$-like function which changes its value from 0 to 1 within
the range $[\xgsb; \xlsb]$ and $q_{1}(x)$ is zero on the boundaries of the $[\xgsb; \xlsb]$ interval and has a peak proportional to the square gradient of the density inside this interval. Thus,
the first two terms in each expression for the resistivity represents a smooth transitions from the gas bulk resistivity to the liquid bulk resistivity, while the third term represents a peak in
the resistivity proportional to the square gradient of the density.

The bulk values $r^{\ell}$ and $r^{g}$ are related to the measurable transport coefficients such as heat conductivity, the diffusion coefficient and the Soret coefficient. We refer for the
details to \cite{glav/grad2}.

Consider a binary mixture of cyclohexane and $n$-hexane, which is the same as in \cite{glav/grad2} and \cite{glav/grad3}. Furthermore we consider a planar interface between liquid and vapor. The
mixture is in a box with gravity directed along axes x from left to right. The gas phase is therefore in the left part of the box and the liquid is in the right part.

We compare the resistivities found in \cite{glav/grad3} to the values obtained from \eqr{eq/Binary/04}. The relative difference between them is almost the same within the range of temperatures
and chemical potential differences considered: $T=\{325,\cdots,335\}$ and $\psi=\{400,\cdots,1000\}$. In \tblr{tbl/Diff/Pc-Ir-4-9-0-3} we give the relative errors for gas and liquid side
coefficients. The data are obtained for $\alpha$-amplitudes, values of which were found to fit best kinetic theory in \cite{glav/grad3}.

\begin{longtable}{l@{\qquad}l@{\qquad}l@{\qquad}l@{\qquad}l@{\qquad}l@{\qquad}l@{\qquad}l@{\qquad}l@{\qquad}l}%
\caption{Relative error in percent between the gas- and liquid- side coefficients obtained by "perturbation cell" and "integral relations" methods at $T_{eq} = 330$ and $\psi_{eq} = 700$ for $\beta = 0.0002$ and $\alpha_{qq} = 9$, $\alpha_{1q} = 0$, $\alpha_{11} = 3$ .} \label{tbl/Diff/Pc-Ir-4-9-0-3}\\%
\hline %
phase & $R_{qq}$     & $R_{11}$  & $R_{22}$  & $R_{q1}$  & $R_{q2}$  & $R_{12}$ \\%
\hline %
gas     & 0.019090   & 0.064642  & 0.058851  & 0.020649  & 0.020680  & 0.097096  \\%
liquid  & 0.019090   & 0.006266  & 0.000432  & 0.036270  & 0.034886  & 6.233983  \\%
\hline %
\end{longtable} %
The relative differences are not more then a few promille. It is larger only for $R_{12}^{\ell}$ which is discussed below.

We also do a consistency check. Consider \eqr{eq/Measurable/01} for two component mixture, which has a form
\begin{equation}\label{eq/Results/01}
\begin{array}{rl}
X_{q} &= R^{\,\prime}_{qq}\,J_{q}^{\,\prime} -  R^{\,\prime}_{q1}\,J_{\xi_{1}} - R^{\,\prime}_{q2}\,J_{\xi_{2}} \\
X_{1} &= R^{\,\prime}_{1q}\,J_{q}^{\,\prime} -  R^{\,\prime}_{11}\,J_{\xi_{1}} - R^{\,\prime}_{12}\,J_{\xi_{2}} \\
X_{2} &= R^{\,\prime}_{2q}\,J_{q}^{\,\prime} -  R^{\,\prime}_{21}\,J_{\xi_{1}} - R^{\,\prime}_{22}\,J_{\xi_{2}} \\
\end{array}
\end{equation}%
The left hand side of each equation must be equal to the right hand side. The difference therefore reflects the error. We give the relative error between the left an the right hand side of
\eqr{eq/Results/01} in percent in \tblr{tbl/X/Pc-Ir-4-9-0-3}. As a testing perturbation we used one of those used in the perturbation cell method.
\begin{longtable}{l@{\qquad}l@{\qquad}l@{\qquad}l@{\qquad}l@{\qquad}l@{\qquad}l@{\qquad}l@{\qquad}l@{\qquad}l}%
\caption{Relative error in percent between the left- and right- hand side of \eqr{eq/Results/01} for coefficients obtained by "perturbation cell" and "integral relations" methods at $T_{eq} = 330$ and $\psi_{eq} = 700$ for $\beta = 0.0002$ and $\alpha_{qq} = 9$, $\alpha_{1q} = 0$, $\alpha_{11} = 3$ .} \label{tbl/X/Pc-Ir-4-9-0-3}\\%
\hline
                & \multicolumn{3}{c}{Integral relations}    & \multicolumn{3}{c}{Perturbation cell} \\
\hline %
phase & $X_{q}$     & $X_{1}$  & $X_{2}$  & $X_{q}$  & $X_{1}$  & $X_{2}$ \\%
\hline %
gas     & 0.059489    & 0.037918  & 0.296959  & 0.046965    & 0.087411  & 0.867098  \\%
liquid  & 0.059489    & 0.172608  & 0.027275  & 0.046851    & 0.216819  & 0.014248  \\%
\hline %
\end{longtable} %
Again, the relative difference is not more then a few promille. Given that this is the case even for a few percent difference in one of the coefficients, we may conclude that the values of the
forces are insensitive to the precise value of this resistivity coefficient. This also indicates that the value of this coefficient obtained in \cite{glav/grad3} has a 6\% error. This does not
necessarily affect, however, the accuracy of the integral relations.

\section{Discussion and conclusions.}\label{sec/Discussion}

In this paper we have derived integral relations for the resistivities to the transport of heat and mass across the interface for mixtures. We have given relations between the local resistivity
profiles and the overall interfacial resistivities.

The integral relations make it possible to calculate the interfacial resistivities in a relatively simple way, using only the equilibrium profiles of the system. This is important especially for
mixtures, for which the computation of a non-equilibrium profiles is much more time consuming.

The integral relations give an insight in the origin of the interfacial resistances. According to \eqr{eq/Binary/03} and \eqr{eq/Binary/04} the interfacial resistivities depend on the variation
of the enthalpy across the interface. The transport coefficients depend on the equilibrium enthalpies which vary a lot through the interface. One can see from the above formulae, that the
dependence on the enthalpy of evaporation (the difference between the enthalpies of the liquid and gas phases) is crucial not only for the diagonal diffusion coefficient, but also for the
off-diagonal coefficients. This is aa important result since cross coefficients are usually neglected in the description of the interfacial phenomena.

Another factor which affects the overall interfacial resistivity is the local resistivity profile. For instance, for the heat resistivity this is the only factor. It is noticeable that the
interfacial resistivity depends on the whole profile of the local resistivity, not only on its bulk values. It is therefore crucial to have complete information about the local resistivity
profiles. We have used sums of a function that smoothly connects the liquid and the vapor values and a peak proportional to the square gradient of the density. In principle one can use any model
for this and further investigation are required. We have shown in \cite{glav/grad3} by comparison with the predictions from kinetic theory, that the local resistivities do have a peak in their
profile and that the overall resistivities therefore depend on the amplitudes of these peaks. Within the current theory these amplitudes are adjustable parameters.

The integral relations are in fact mathematical equalities. Given the local resistivities defined through the local force-flux relations, for instance $r^{e}$ from \eqr{eq/Resistivities/01},
\eqr{eq/Resistivities/02} follows and one can consider \eqr{eq/Binary/03} as a definition of the overall interfacial resistivities $R^{e}$ used in \eqr{eq/Resistivities/02}. It means that the
force-flux relations \eqref{eq/Measurable/01} for the whole surface follow from the local force-flux relations. One therefore does not need the excess entropy production
\eqr{eq/ExcessEntropy/10} to obtain \eqr{eq/Resistivities/02}.
%This fact is independent from the original reasons to write the phenomenological relations based on the expression for

This allows us to use them as a test for the accuracy of the numerical solution of the non-equilibrium case. Given that both local and overall linear laws are true independently, the different
methods to obtain the overall resistivities give information about the accuracy of the method. The discussion below \tblr{tbl/Diff/Pc-Ir-4-9-0-3} and \tblr{tbl/X/Pc-Ir-4-9-0-3} is based on this
observation.

\appendix

\section{Local resistivities.}\label{sec/Appendix/Resistivities}

We need to relate the resistivities $r$ from \eqr{eq/Resistivities/01} to the resistivities $p$ from \eqr{eq/Measurable/04}. This is done by comparing the coefficients at the same fluxes in
these equations. To do this we need to translate the set of fluxes used in \eqr{eq/Measurable/04}, $\{J_{q}^{\,\prime},\,J_{1},\cdots,J_{n-1}\}$, to the set of fluxes used in
\eqr{eq/Resistivities/01}, $\{J_{e},\,J_{\xi_{1}},\cdots,J_{\xi_{n}}\}$. This is done with the help of the relation
\begin{equation}\label{eq/Appendix/Resistivities/01}
\begin{array}{rl}
J_{i}               =& \displaystyle J_{\xi_{i}} - \xi_{i}\sum_{k=1}^{n}{J_{\xi_{k}}} \\
J_{q}^{\,\prime}    =& \displaystyle J_{e} - \sum_{k=1}^{n}{\widetilde{h}_{k}J_{\xi_{k}}} \\
\end{array}
\end{equation}
Substituting $J_{q}^{\,\prime}$ and $J_{i}$ into the first of \eqr{eq/Measurable/04} we obtain
\begin{equation}\label{eq/Appendix/Resistivities/02}
\displaystyle \nabla_{\perp}\frac{1}{T} = \displaystyle r^{\,\prime}_{qq}\,J_{e} - \sum_{i=1}^{n-1}{J_{\xi_{i}}\Big(r^{\,\prime}_{qq}\widetilde{h}_{i} + r^{\,\prime}_{qi} - \sum_{i=k}^{n-1}{r^{\,\prime}_{qk}\xi_{k}}\Big)}  - J_{\xi_{n}}\Big(r^{\,\prime}_{qq}\widetilde{h}_{n} - \sum_{i=k}^{n-1}{r^{\,\prime}_{qk}\xi_{k}}\Big)\\
\end{equation}%
Comparing it with the first of \eqr{eq/Resistivities/01} we obtain
\begin{equation}\label{eq/Appendix/Resistivities/03}
\begin{array}{rl}
r^{e}_{qq} =& \displaystyle r^{\,\prime}_{qq} \\
r^{e}_{qi} =& \displaystyle r^{\,\prime}_{qq}\widetilde{h}_{i} - \sum_{k=1}^{n-1}{r^{\,\prime}_{qk}\,\xi_{k}} + r^{\,\prime}_{qi},\quad i=\nmorange \\
r^{e}_{en} =& \displaystyle r^{\,\prime}_{qq}\widetilde{h}_{n} - \sum_{k=1}^{n-1}{r^{\,\prime}_{qk}\,\xi_{k}} \\
\end{array}
\end{equation}
which are the first 3 equations of \eqr{eq/Measurable/06}.

In order to obtain the remaining relations we consider the second of \eqr{eq/Resistivities/01}, which gives
\begin{equation}\label{eq/Appendix/Resistivities/04}
\begin{array}{rl}
\displaystyle \sum_{j=1}^{n}{\xi_{j}\,\nabla_{\perp}\frac{\tilde{\mu}_{j}}{T}} &= \displaystyle J_{e}\sum_{j=1}^{n}{r^{e}_{jq}\xi_{j}} - \sum_{i=1}^{n}{J_{\xi_{i}}\sum_{j=1}^{n}{r^{e}_{ji}\xi_{j}}} \\
\displaystyle \nabla_{\perp}\frac{\tilde{\mu}_{j}}{T}-\nabla_{\perp}\frac{\tilde{\mu}_{n}}{T} &= \displaystyle J_{e}(r^{e}_{jq}-r^{e}_{ne}) -
\sum_{i=1}^{n}{J_{\xi_{i}}(r^{e}_{ji}-r^{e}_{ni})},\quad j=\nmorange
\end{array}
\end{equation}%

Furthermore we use \eqr{eq/ExcessEntropy/04a} which in case of the transport in the direction only perpendicular to the surface becomes
\begin{equation}\label{eq/Appendix/Resistivities/05}
\sum_{i=1}^{n}{\xi_{i}\left(\nabla_{\perp}\frac{\widetilde{\mu}_{i}}{T} - \widetilde{h}_{i}\nabla_{\perp}\frac{1}{T}\right)} = 0
\end{equation}%
Together with the second of \eqr{eq/Measurable/04} it gives
\begin{equation}\label{eq/Appendix/Resistivities/06}
\begin{array}{rl}
\displaystyle \sum_{i=1}^{n}{\xi_{i}\,\nabla_{\perp}\frac{\widetilde{\mu}_{i}}{T}} &= \displaystyle \sum_{i=1}^{n}{\xi_{i}\,\widetilde{h}_{i}\,\nabla_{\perp}\frac{1}{T}}\\
\displaystyle \nabla_{\perp}\frac{\psi_{i}}{T} &= \displaystyle \eta_{i}\nabla_{\perp}\frac{1}{T} + r^{\,\prime}_{jq}\,J_{q}^{\,\prime} - \sum_{i=1}^{n-1}{r^{\,\prime}_{ji}\,J_{i}}
\end{array}
\end{equation}
Substituting $\nabla_{\perp}(1/T)$ from \eqr{eq/Appendix/Resistivities/02} and $J_{q}^{\,\prime}$ and $J_{i}$ from \eqr{eq/Appendix/Resistivities/01} we obtain the left hand size of
\eqr{eq/Appendix/Resistivities/06} expressed in terms of the fluxes $J_{e}$ and $J_{\xi_{1}}$ and the resistivities $r^{\,\prime}$. Comparing the result with \eqr{eq/Appendix/Resistivities/04}
we obtain the following equations sets
\begin{subequations}\label{eq/Appendix/Resistivities/07}
\begin{equation}\label{eq/Appendix/Resistivities/07a}
\begin{array}{rl}
\sum_{j=1}^{n}{r^{e}_{jq}\xi_{j}}  &= r^{\,\prime}_{qq}\sum_{k=1}^{n}{\xi_{k}\,\widetilde{h}_{k}}\\\\
r^{e}_{jq} - r^{e}_{ne} &= r^{\,\prime}_{qq}\eta_{j} + r^{\,\prime}_{jq}, \quad j=\nmorange\\
\end{array}
\end{equation}
\begin{equation}\label{eq/Appendix/Resistivities/07b}
\begin{array}{rl}
\sum_{j=1}^{n}{r^{e}_{jn}\xi_{j}}  &= (r^{\,\prime}_{qq}\widetilde{h}_{n} -\sum_{k=1}^{n-1}{r^{\,\prime}_{qk}\,\xi_{k}})\sum_{k=1}^{n}{\xi_{k}\,\widetilde{h}_{k}}\\\\
r^{e}_{jn} - r^{e}_{nn} &= (r^{\,\prime}_{qq}\eta_{j}+r^{\,\prime}_{jq})\widetilde{h}_{n} -\sum_{k=1}^{n-1}{(r^{\,\prime}_{qk}\,\eta_{j}+r^{\,\prime}_{jk})\,\xi_{k}}, \quad j=\nmorange\\
\end{array}
\end{equation}
\begin{equation}\label{eq/Appendix/Resistivities/07c}
\begin{array}{rl}
\sum_{j=1}^{n}{r^{e}_{ji}\xi_{j}}  &= (r^{\,\prime}_{qq}\widetilde{h}_{i} - \sum_{k=1}^{n-1}{r^{\,\prime}_{qk}\,\xi_{k}} + r^{\,\prime}_{qi}) \sum_{k=1}^{n}{\xi_{k}\,\widetilde{h}_{k}}, \quad i=\nmorange\\\\
r^{e}_{ji} - r^{e}_{ni} &= (r^{\,\prime}_{qq}\eta_{j}+r^{\,\prime}_{jq})\widetilde{h}_{i} -\sum_{k=1}^{n-1}{(r^{\,\prime}_{qk}\,\eta_{j}+r^{\,\prime}_{jk})\,\xi_{k}} + (r^{\,\prime}_{qi}\,\eta_{j}+r^{\,\prime}_{ji}), \quad j=\nmorange, i=\nmorange\\
\end{array}
\end{equation}
\end{subequations}
solving which we obtain the relations between the remaining resistivities
\begin{equation}\label{eq/Appendix/Resistivities/08}
\begin{array}{rl}
r^{e}_{jq} =& \displaystyle r^{\,\prime}_{qq}\widetilde{h}_{j} - \sum_{k=1}^{n-1}{r^{\,\prime}_{ke}\,\xi_{k}} + r^{\,\prime}_{jq},\quad j=\nmorange \\
r^{e}_{ne} =& \displaystyle r^{\,\prime}_{qq}\widetilde{h}_{n} - \sum_{k=1}^{n-1}{r^{\,\prime}_{ke}\,\xi_{k}} \\
r^{e}_{ji} =& \displaystyle r^{\,\prime}_{qq}\widetilde{h}_{j}\widetilde{h}_{i} - \sum_{k=1}^{n-1}{\xi_{k}(r^{\,\prime}_{ke}\widetilde{h}_{i}+r^{\,\prime}_{qk}\widetilde{h}_{j})} + r^{\,\prime}_{jq}\widetilde{h}_{i}+r^{\,\prime}_{qi}\widetilde{h}_{j}+ \\
        & \displaystyle + r^{\,\prime}_{ji} - \sum_{k=1}^{n-1}{r^{\,\prime}_{ki}\,\xi_{k}} - \sum_{k=1}^{n-1}{r^{\,\prime}_{jk}\,\xi_{k}} + \sum_{k=1}^{n-1}\sum_{l=1}^{n-1}{r^{\,\prime}_{kl}\,\xi_{k}\,\xi_{l}},\quad j,i=\nmorange \\
r^{e}_{jn} =& \displaystyle r^{\,\prime}_{qq}\widetilde{h}_{j}\widetilde{h}_{n} - \sum_{k=1}^{n-1}{\xi_{k}(r^{\,\prime}_{ke}\widetilde{h}_{n}+r^{\,\prime}_{qk}\widetilde{h}_{j})} + r^{\,\prime}_{jq}\widetilde{h}_{n} - \sum_{k=1}^{n-1}{r^{\,\prime}_{jk}\,\xi_{k}} + \sum_{k=1}^{n-1}\sum_{l=1}^{n-1}{r^{\,\prime}_{kl}\,\xi_{k}\,\xi_{l}},\quad j=\nmorange \\
r^{e}_{ni} =& \displaystyle r^{\,\prime}_{qq}\widetilde{h}_{n}\widetilde{h}_{i} - \sum_{k=1}^{n-1}{\xi_{k}(r^{\,\prime}_{ke}\widetilde{h}_{i}+r^{\,\prime}_{qk}\widetilde{h}_{n})} + r^{\,\prime}_{qi}\widetilde{h}_{n} - \sum_{k=1}^{n-1}{r^{\,\prime}_{ki}\,\xi_{k}} + \sum_{k=1}^{n-1}\sum_{l=1}^{n-1}{r^{\,\prime}_{kl}\,\xi_{k}\,\xi_{l}},\quad i=\nmorange \\
r^{e}_{nn} =& \displaystyle r^{\,\prime}_{qq}\widetilde{h}_{n}^{2}  - \widetilde{h}_{n}\sum_{k=1}^{n-1}{\xi_{k}(r^{\,\prime}_{ke}+r^{\,\prime}_{qk})} + \sum_{k=1}^{n-1}\sum_{l=1}^{n-1}{r^{\,\prime}_{kl}\,\xi_{k}\,\xi_{l}} \\
\end{array}
\end{equation}

As one can confirm the symmetry of the $r^{\,\prime}$-matrix leads to the symmetry of the $r^{e}$-matrix and vice versa. We therefore do not give the expressions for $r^{e}_{jq}$, $r^{e}_{ne}$
and $r^{e}_{jn}$ in \eqr{eq/Measurable/06}.

The relations \eqref{eq/Measurable/07} between the $r^{e}$- and $r$- resistivities are derived in the similar manner.

%Included for Gather Purpose only:
%input "vdWbib.bib"
\bibliographystyle{unsrt}
%\bibliography{vdWbib,interface}

\end{document}